\documentstyle[12pt]{article}
\newcommand{\be}{\begin{equation}}
\newcommand{\ee}{\end{equation}}
\newcommand{\bea}{\begin{eqnarray}}
\newcommand{\eea}{\end{eqnarray}}
\newcommand{\ba}{\begin{array}}
\newcommand{\ea}{\end{array}}
\newcommand{\bd}{\begin{description}}
\newcommand{\ed}{\end{description}}
\newcommand{\ds}{\displaystyle}

\def\sym#1#2#3#4#5#6{\left(\matrix{#1&#2\cr#3&#4\cr}\right|
                     \left.\matrix{#5\cr#6\cr}\right)}

\def\inbar{\vrule height1.5ex width.4pt depth0pt}
\def\IB{\relax{\rm I\kern-.18em B}}
\def\IC{\relax\,\hbox{$\inbar\kern-.3em{\rm C}$}}
\def\ID{\relax{\rm I\kern-.18em D}}
\def\IE{\relax{\rm I\kern-.18em E}}
\def\IF{\relax{\rm I\kern-.18em F}}
\def\IG{\relax\,\hbox{$\inbar\kern-.3em{\rm G}$}}
\def\IH{\relax{\rm I\kern-.18em H}}
\def\II{\relax{\rm I\kern-.18em I}}
\def\IK{\relax{\rm I\kern-.18em K}}
\def\IL{\relax{\rm I\kern-.18em L}}
\def\IM{\relax{\rm I\kern-.18em M}}
\def\IN{\relax{\rm I\kern-.18em N}}
\def\IO{\relax\,\hbox{$\inbar\kern-.3em{\rm O}$}}
\def\IP{\relax{\rm I\kern-.18em P}}
\def\IQ{\relax\,\hbox{$\inbar\kern-.3em{\rm Q}$}}
\def\IR{\relax{\rm I\kern-.18em R}}
\def\ZZ{\relax{\sf Z\kern-.4em Z}}

\def\figskip{\phantom{xxxxxxxxxxxxxxxxxxxxxxxxxxxxxxxxxxxxxxxxxxxxxxx}}

\def\us#1{{\underline{#1}}}

\newcommand{\sect}[1]{ \section{#1} \setcounter{equation}{0} }

\def\npb#1(#2)#3{{ Nucl. Phys. }{\bf B#1} (#2) #3}
\def\plb#1(#2)#3{{ Phys. Lett. }{\bf #1B} (#2) #3}
\def\pla#1(#2)#3{{ Phys. Lett. }{\bf #1A} (#2) #3}
\def\mpla#1(#2)#3{{ Mod. Phys. Lett. }{\bf A#1} (#2) #3}
\def\ijmpa#1(#2)#3{{Int. J. Mod. Phys. }{\bf A#1} (#2) #3}
\def\cmp#1(#2)#3{{ Comm. Math. Phys. }{\bf #1} (#2) #3}
\def\cqg#1(#2)#3{{ Class. Quantum Grav. }{\bf #1} (#2) #3}
\def\jmp#1(#2)#3{{ J. Math. Phys. }{\bf #1} (#2) #3}
\def\anp#1(#2)#3{{ Ann. Phys. }{\bf #1} (#2) #3}
\def\mm{{\sl Essays on Mirror Manifolds} (Ed. S.-T. Yau),
         Int. Press. Co., Hong Kong, (1992)}

\def\a{\alpha}

\def\p1{\phi_1}
\def\p2{\phi_2}

\def\({\lbrack}
\def\){\rbrack}

\begin{document}

\title{Mirror Maps and Instanton
Sums for Complete Intersections
in Weighted Projective Space\thanks{Supported by the Deutsche
Forschungsgemeinschaft and the EC under contract SC1-CT92-0789}}

\author{ {Albrecht Klemm}\thanks{email:aklemm@physik.tu-muenchen.de}
and {Stefan Theisen}\thanks{email:sjt@dmumpiwh} \\
{Sektion Physik} \\
{Ludwig-Maximilians Universit\"at M\"unchen} \\
{Theresienstra\ss e 37} \\
{8000 M\"unchen 40, Federal Republic of Germany} \\ }
\date{April 1993}
\maketitle
\begin{picture}(5,2.5)(-320,-360)
\put(12,-90){LMU-TPW 93-08}
\end{picture}

\begin{abstract} We consider a class of Calabi-Yau
compactifications which are constructed as a complete intersection in
weighted projective space. For manifolds with one K\"ahler modulus we
construct the mirror manifolds and calculate the instanton sum.
\end{abstract}

\sect{Introduction}
When considering symmetric $(2,2)$ super-conformal
field theories as internal conformal field theories
relevant for string theory, one immediately
observes that due to the arbitrariness in the
assignment of the relative sign of the two $U(1)$
charges, there is a complete symmetry between
two theories for which the $(a,c)$ and the $(c,c)$
rings are interchanged\cite{Dixon}\cite{LVW}.
On the geometrical level,
when considering conformal field theories which admit
an interpretation as compactification on Calabi-Yau
manifolds, the fields in the two rings correspond to the
K\"ahler and the complex structure moduli of the
Calabi-Yau space, respectively. The observation
on the conformal field theory level mentioned above,
then lead to the so-called
mirror hypothesis. In rough terms, it states that
for each Calabi-Yau manifold there should exist
one with the two hodge numbers
$h_{1,1}$ and $h_{2,1}$ interchanged and that
string propagation on these two manifolds should
be identical (for further details on mirror symmetry, see \cite{MM}).
The special case of manifolds with $h_{2,1}=0$ is discussed in
\cite{CDP}.

This mirror hypothesis has to date not been proven yet,
but looking at the Hodge numbers of Calabi-Yau
spaces constructed so far, one observes a rough
symmetry and for some cases the mirror manifolds have
been constructed. Here we add some more examples:
manifolds with $h_{1,1}=1$ which are given as complete
intersections in weighted projective space.

Aside from being interesting from a mathematical
point of view, mirror symmetry can also be put
to practical use for the computation of certain
physical quantities of the low energy effective field theory
such as the Yukawa couplings of charged matter fields
or the K\"ahler metric for the moduli fields
\cite{CDGP}\cite{Morrison}\cite{KT1}\cite{Font}.
This is so since whereas the Yukawa couplings of
three ${\bf 27}$'s fields,
which are related to the $(2,1)$ moduli fields by
$n=2$ superconformal symmetry, receive no $\sigma$-model
corrections \cite{DG} (perturbatively and non-perturbatively),
the couplings of three $\overline{\bf 27}$'s, related to the $(1,1)$ moduli,
do receive corrections from world sheet instantons and are thus (probably)
impossible to compute directly. The former couplings
depend on the complex structure moduli, the latter
are functions of the K\"ahler moduli and are  given by
$$
\kappa_{ijk}=\int_X b^{(1,1)}_i\wedge b^{(1,1)}_j\wedge b^{(1,1)}_k
+\sum_{m,n}m^{-3}e^{-\int_{I_{n,m}}\imath^*(J)}
\int_{I_{m,n}}\imath^*(b^{(1,1)}_i)\int_{I_{m,n}}\imath^*(b^{(1,1)}_j)
\int_{I_{m,n}}\imath^*(b^{(1,1)}_k)\ .
$$
Here, $b^{(1,1)}_i$ are $(1,1)$ forms and $I_{m,n}$ is the
$m$-fold cover of a rational curve on $X$ of degree $n$,
$\imath$ its inclusion $I_{m,n}\hookrightarrow X$. The K\"ahlerform
$J=\sum_i^{h_{1,1}} t_i b^{(1,1)}_i$ is expanded in the basis of $H^3(X,\ZZ)$
s.t. the moduli $t_i$ parameterize the complexified K\"ahler cone.
The first term is the sigma model tree level contribution and the
second term the contribution from world-sheet
instantons.

The mirror hypothesis now amounts to the identification
of this coupling with the corresponding coupling
on the mirror manifold $X^\prime$, which is a function of the
$(2,1)$ moduli $t^\prime_i$ of which there is an equal
number as there are $(1,1)$ moduli on the original manifold.
The relation between
$(b^{(1,1)}_i(X),t_i)\leftrightarrow (b^{(2,1)}_i(X^\prime),t_i^\prime)$
is described by the mirror map. The ${\bf 27}^3$ couplings on $X^\prime$
can be computed, at least in principle, from
knowledge of the periods of the holomorphic three-form,
i.e. the solutions of the Picard-Fuchs equations,
and special geometry. We will extend the list of models
which have been treated along these lines in refs.
\cite{CDGP}\cite{Morrison}\cite{Font}\cite{KT1}\cite{LT}
to models with $h_{(1,1)}=1$ which are given as complete
intersections in weighted projective space.

\sect{The manifolds and their mirrors}
There exist examples of complete intersections with $c_1=0$ in ordinary
projective
complex space: two cubics in $\IP^5$, a quartic
and a quadric in $\IP^5$,
two quadrics and a cubic in $\IP^6$ and four quadrics in $\IP^7$.
Passing to weighted
projective spaces as ambient space, several thousand Calabi-Yau manifolds
can be
constructed \cite{DK}.
As in the hypersurface case \cite{Ks}\cite{ks},
the  complete intersections inherit, in most of the cases, cyclic quotient
singularities from the ambient space and
canonical desingularizations are
required to obtain a smooth manifold.
As the process of the canonical desingularization introduces
irreducible exceptional divisors, the dimension of the fourth homology
group and hence the number of (1,1)-forms or K\"ahler
moduli for these examples is usually bigger than one.
Within the class under consideration we have \cite{DK}
only the following five families\footnote{The list in
II.4.6. of \cite{Fletcher} is not complete.} listed in Table 1, which are,
as the intersection locus avoids all singularities
of the ambient space, one K\"ahler modulus examples.
The Hodge diamond can be calculated by the general formulas of \cite{DK},
which automatically take care of the desingularizations,
but as we deal here
only with smooth intersections, we will use the somewhat simpler
adjunction formula to get the Euler characteristic
of the $(m-1)-k$-dimensional smooth
complete intersection $X:=X_{\{d_1,\ldots,d_k\}}\in\IP^{m-1}(\underline w)$
of $k$ polynomial constraints
$p_1 = \ldots = p_k=0$ in $\IP^{m-1}(\underline w)$ of degree
$d_1,\ldots,d_k$ as the coefficient of $J^{m-1}$ in the
formal expansion in $J$ of the quotient
\be
{\prod_{i=1}^{m} (1+w_{i} J)\prod_{j=1}^k d_j J\over
\prod_{j=1}^k(1+ d_j J)\prod_{i=1}^m w_i}.
\label{eulform}
\ee

\begin{table}[h]
\begin{center}
\begin{tabular}{|r|r|r|r|r|} \hline
\multicolumn{1}{|c|} {$N^0$}
&\multicolumn{1}{|c|}{$X=X_{(d_1,d_2) }\subset \IP^5(\us w)$}
&\multicolumn{1}{|c|}{$\chi(X)$}
&\multicolumn{1}{|c|}{$h^{1,2}(X)$}
&\multicolumn{1}{|c|}{$h^{1,1}(X)$}
 \\[.31 mm] \hline \hline &&&&\\[-3.5 mm]
 1 &$ X_{(4,4) }\subset\IP^5(1,1,2,1,1,2)  $&   -144  &  73 & 1  \\
 2 &$ X_{(6,6) }\subset\IP^5(1,2,3,1,2,3)  $&   -120  &  61 & 1  \\
 3 &$ X_{(4,3) }\subset\IP^5(2,1,1,1,1,1)   $&  -156  &  79 & 1  \\
 4 &$ X_{(6,2) }\subset\IP^5(3,1,1,1,1,1)   $&  -256  & 129 & 1  \\
 5 &$ X_{(6,4) }\subset\IP^5(3,2,2,1,1,1)   $&  -156  &  79 & 1  \\
 \hline
\end{tabular}
\caption{ Complete intersections in weighted projective
space with one K\"ahler modulus}
\end{center}
\end{table}

To construct the mirror manifold we first have to specify a
configuration of the polynomial constraints, which is transversal
in $\IP^{m-1}(\underline w)$ for almost all values of
the complex deformation parameters.
This means that all $k \times k$ subdeterminants of
$\left({\partial p_i\over \partial x_j}\right)$ are allowed
to  vanish for generic values the deformation parameter
only at $x_1=\ldots =x_m=0$.
As in \cite{Fletcher} one can use Bertinis Theorem 
to formulate a transversality criterium as the requirement
that certain monomials occurs in the polynomials. Let
$J=\{j_1,\ldots,j_{|J|}\}\in \{1,\ldots,m\}$ be an index set and
$X^{M_J}_J:=x_{j_1}^{m_1}\ldots x_{j_{|J|}}^{m_{|J|}}$
(where $m_j\in \IN_0$)
a monomial. For $k=2$ (see \cite{Fletcher} Theorem I.5.7 and \cite{DK}
for generalisations) transversality requires for all index sets
$J$ that there occur
\begin{itemize}

\item[(a)] either a monomial $X^{M^1_J}_J$ in $p_1$ and
a monomial $X^{M^2_J}_J$ in $p_2$

\item[(b)] or a monomial $X^{M^1_J}_J$ in $p_1$ and
$|J|-1$ monomials $X^{M^i_J}_J x_{e_i}$ in $p_2$ with
distinct $e_i$'s

\item[(c)] or (b) with $p_1$ and $p_2$ interchanged

\item[(d)] or $|J|$ monomials $X^{M^i_J}_J x_{e^1_i}$ in $p_1$ with
distinct $e^1_i$'s and $|J|$ monomials $X^{M^j_J}_J x_{e^2_j}$
in $p_2$ with distinct $e^2_j$'s s.t. $\{e^1_i,e^2_j\}$
contains at least $|J|+1$ distinct elements
\end{itemize}

We then search for a discrete isomorphy group
$G$ which leaves the holomorphic $(3,0)$ form (see (\ref{30form}) below)
invariant.
In the cases where we have succeeded in constructing the mirror,
$G$ is Abelian and can be represented by phase multiplication
\be
x_i \mapsto \exp{(2\pi i g_i)} \, x_i\quad i=1,\ldots,m
\label{groupaction}
\ee
$g_i\in \IQ$ on the homogeneous $\IP^{m-1}(\underline w)$
coordinates. The singular orbifold
$X/G$ has precisely the same type of Gorenstein
singularities which one encounters in the cases of complete
intersections or hypersurfaces in weighted projective
spaces and the canonical
desingularization $X^\prime=\widehat{ X/G}$ has the ``mirror'' Hodge
diamond with $h_{i,j}(\widehat{ X/G})=h_{i,3-j}(X)$.

Let us give the explicit construction for the two quartics
$X_{(4,4)}$ in $\IP^5(1,1,2,1,1,2)$.
As the one parameter family we choose
the following form of the polynomial constraints
(comp. \cite{LT})
\be
p_1=x_1^4+x_2^4+2 x_3^2 - 4 \, \alpha \, x_4 x_5 x_6=0,\qquad
p_2=x_4^4+x_5^4+2 x_6^2 - 4 \, \alpha \, x_1 x_2 x_3=0,
\label{cy1}
\ee
which is transversal for almost all
values of the deformation parameter $\alpha$, except $\alpha=0$,
$\alpha^8=1$ or $\alpha=\infty$.

The maximal symmetry group $G$ is a $\ZZ_2\times \ZZ_2\times \ZZ_{16}$
and we may choose the following generators:
$$
g^{(1)}=\ds{{1\over 2}\, (0,1,1,0,0,0)},\qquad
g^{(2)}=\ds{{1\over 2}\, (0,0,0,0,1,1)},\qquad
g^{(3)}=\ds{{1\over 16}\, (0,4,0,1,13,2)}.
$$
The group action is understood as in (\ref{groupaction}).
$G$ has four $\ZZ_2$ subgroups,
which leave four curves
$C^{(-24)}_1,C^{(-24)}_2,C^{(-24)}_4,C^{(-24)}_5$ of type
$X_{(4,4)}(1,1,1,2)$ invariant.
They are specified by the subsets of $X$ with $(x_2=x_3=0)$,
$(x_1=x_3=0)$,  $(x_4=x_6=0)$ and $(x_5=x_6=0)$ respectively and we may
calculate their Euler characteristic (indicated as right upper index) by
(\ref{eulform}).
The $\ZZ_2$ acts on the coordinates of the fibers of the
normal bundle to the curves by $g={1\over 2}(1,1)$
and the canonical desingularization of the $A_1$-type $\IC^2/\ZZ_2$
singularity in the fibers adds one exceptional divisor per curve.
Moreover we have two $\ZZ_4$ subgroups of $G$ leaving the curves
$C^{(-8)}_3$ $(x_1=x_2=0)$ and $C_6^{(-8)}$ $(x_4=x_5=0)$ of type
$X_{(4,4)}(1,1,2,2)$ invariant. The corresponding
action on the coordinates of the normal fiber is of type
$g={1\over 4}(1,3)$ and gives rise to a $A_3$-type singularity,
which adds, by canonical desingularizaton,
three exceptional divisors per curve.
So we expect $10$ irreducible exceptional divisors from desingularization
of the fixed curves.
Note that the curves $C_1,C_2,C_3$ have triple intersections
in eight points, which we group into the following fixed point sets
$P^{(2)}_{1,2}$ $(x_{1,2}=x_4=x_5=x_6=0)$,
$P^{(4)}_3$ $(x_3=x_4=x_5=x_6=0)$,
where the multiplicity of the points in $X$ is again indicated as left
upper index. It can also be obtained from (\ref{eulform}) as
a point has Euler number one.
Obviously to
get their multiplicity on the orbifold $X/G$, we have to multiply by
a factor $|I|/|G|$ where $|I|$ and $|G|$ are the orders of the isotropy
group of the set and of $G$, respectively.
Analogous remarks apply to the intersection of
$C_4,C_5,C_6$ and $P^{(2)}_{4,5}$ $(x_{4,5}=x_1=x_2=x_3=0)$,
$P^{(4)}_6$ $(x_6=x_1=x_2=x_3=0)$. The schematic intersection pattern
of the fixed set singularities is depicted in Fig. 1.

\figskip
\vskip  -0.8  mm
\includegraphics{f1.ps}
\setlength{\unitlength}{ 10 mm}
\vskip  4 mm
\begin{picture}(16,5)(0,0)

\put(1.55,1.9){$P_1$}
\put(3.15,1.9){$P_2$}
\put(4.8,1.9){$P_3$}
\put(5.4,1.4){$\ZZ_2:C^{(-24)}_2$}
\put(5.4,3.2){$\ZZ_2:C^{(-24)}_1$}
\put(6.3,2.4){$\ZZ_4:C^{(-8)}_3$}
\put(4.4,4){$\ZZ_2\! \times \! \ZZ_8$}
\put(2.75,4){$\ZZ_2\! \times\!  \ZZ_{16}$}
\put(1.05,4){$\ZZ_2\!  \times\!  \ZZ_{16}$}

\put(10.05,1.9){$P_4$}
\put(11.65,1.9){$P_5$}
\put(13.3,1.9){$P_6$}
\put(13.9,1.4){$\ZZ_2:C^{(-24)}_5$}
\put(13.9,3.2){$\ZZ_2:C^{(-24)}_4$}
\put(14.8,2.4){$\ZZ_4:C^{(-8)}_6$}
\put(12.9,4){$\ZZ_2\! \times \! \ZZ_8$}
\put(11.25,4){$\ZZ_2\! \times\!  \ZZ_{16}$}
\put(9.55,4){$\ZZ_2\!  \times\!  \ZZ_{16}$}

\put(7.4,.5){\bf Fig. 1}
\end{picture}

The isotropy group $I$ of the points
$P_{1,2}$ and $P_{4,5}$ is $\ZZ_2 \times \ZZ_{16}$
whose generators in a local
coordinate system $(x_1,x_2,x_3)$ with the
fixed point as origin may be represented as
$$g^{(1)}={1\over 2}(0,1,1),\qquad g^{(2)}={1\over 16}(1,5,10).$$
That is locally, we have an Abelian $\IC^3/(\ZZ_2\times \ZZ_{16})$
singularity, whose
canonical desingularisation can be constructed by toric geometry
\cite{MOP},\cite{Oda},\cite{Roan1},\cite{DK}.
The $\ZZ_2\times \ZZ_8$ isotropy group $I$ of the points $P_3,P_6$ is
generated by
$$g^{(1)}={1\over 2} (0,1,1),\qquad g^{(2)}={1\over 8} (1,2,5).$$
We show in figure (2 a) and (2 b) the side $\Delta$ (trace) opposite
to the apex of the three
dimensional simplicial fan which, together with the
lattice $\Lambda$, describes the
topological data of the local desingularization processes for the
two types of fixed points. The points are the intersection of
$\Lambda$ with $\Delta$.
Their location is given (see eg. \cite{Roan1}) by
\be{\cal P}=\left\{\sum_{i=1}^3 \vec e_i g_i\,\Bigg|\,(g_1,g_2,g_3)\in
\IQ^3,\left(\ba{rrr}
e^{2 \pi i g_1} &&\cr
&e^{2 \pi i g_2} &\cr
&&e^{2 \pi i g_3}\cr\ea\right)\in I,
\sum_{i=1}^3 g_i=1,g_i\ge 0\right\},\label{exc}
\ee
where the vectors $\vec e_1,\vec e_2,\vec e_3$
span the equilateral triangle from its center.
According to the general theory \cite{Oda,Roan1},
points in the inside of the triangle correspond to
new exceptional divisors while points on the sides of the triangles
correspond to exceptional divisors which are also present
over the generic points of the curves,
which intersect in the fixed point. Counting the inner
points in Fig (2 a,b) we see that each desingularization of the
four $\ZZ_2\times \ZZ_{16}$ orbifold points adds $13$ new
exceptional divisors, while each desingularization of the
two $\ZZ_2\times \ZZ_{8}$ orbifold singularities adds $5$
new exceptional divisors.
Together with the ones over the curves we have $72$ exceptional
divisors from
the desingularization and adding the pullback of the K\"ahler form
of the ambient space we obtain $h_{1,1}(\widehat{X/G)}=73$.

\vskip  -8  mm
\includegraphics{f2.ps}
\includegraphics{f3.ps}
\setlength{\unitlength}{ 10 mm}
\figskip
\vskip  6 mm
\begin{picture}(16,7)(0,0)
\put(0.3,1.2){$\vec e_1$}
\put(6.4,1.2){$\vec e_2$}
\put(3.5,6.3){$\vec e_3$}
\put(9.1,1.2){$\vec e_1$}
\put(15.2,1.2){$\vec e_2$}
\put(12.1,6.4){$\vec e_3$}
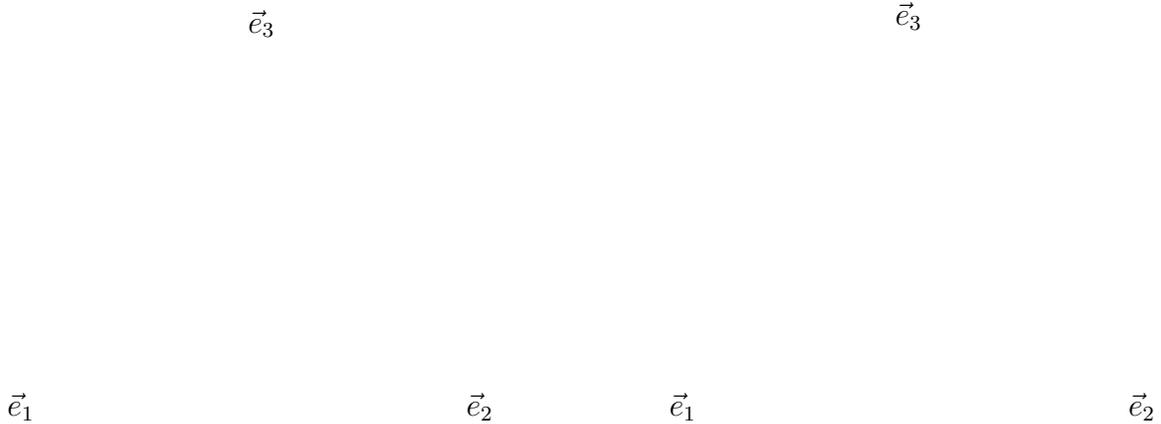
\put(0.5,0) {{\bf Fig. 2 a)} Trace of the $\ZZ_2\times \ZZ_{16}$ fan.}
\put(9.4,0){{\bf Fig. 2 b)} Trace of the $\ZZ_2\times \ZZ_{8}$ fan.}
\end{picture}
\vskip  2 mm

The Euler characteristic $\chi(\widehat{X/G})$ is most easily calculated
by the ``orbifold'' formula of \cite{DHVW}, which for the Abelian case
simplifies to
\be
\chi(\widehat{X/G})={1\over |G|}
\left(\chi(X)-\sum_{I} \chi(S_I)\right)
+\sum_{I} {|I|^2 \over |G|}\chi(S_I),
\label{orbiform}
\ee
where the sum is over all subsets $S_I$ of $X$ which are fixed under the
isotropy groups $I\in G$. Application to the case at hand yields
$\chi(\widehat{X/G})={1\over 64}\left(-144-2\cdot 8 - 4\cdot (-24-8)
-2\cdot (-8-8) \right)+ {8 \cdot 16^2\over 64} + {8 \cdot 32^2\over 64}
+ {4\cdot (-32) \cdot 2^2\over 64}+{2\cdot(-16)\cdot 4^2\over 64}=144,$
hence we have indeed constructed a mirror configuration.

On the following transversal one parameter family
$p_1=x_1^6+2x_2^3+3x_3^2-6\, \alpha\, x_4 x_5 x_6=0$,
$p_2=x_4^6+2x_5^3+3x_6^2 - 6 \, \alpha \, x_1 x_2 x_3=0$
of the second example, we have a group action
$g={1\over 36}\, (6,0,0,1,14,21)$, which has two
fixed curves  of order two, $C^{(-12)}_{1,3}$
($A_1$-type), and two of order
three, $C^{(-6)}_{2,4}$  ($A_2$-type).
$C^{(-12)}_{1}$ and $C^{(-6)}_{2}$ intersect
in six points namely three of order $12$, two of order $18$ and one of
order $36$, which have the following generators for their isotropy
groups:
${1\over 12}\, (1,2,9)$, ${1\over 18}\, (1,14,3)$ and
${1\over 36}\, (1,14,21)$.
An analogeous pattern occurs for $C^{(-12)}_{2}$ and $C^{(-6)}_{4}$.
Applying (\ref{exc}) and (\ref{orbiform}) the reader may check as above
that the resolved $\widehat{X/G}$ has the mirror Hodge diamond.

\sect{The Picard-Fuchs equations and their solutions}
Since the dimension of the third cohomology of the manifolds
$X^\prime$ considered here is four, there must
exist a linear relation between the holomorphic three-form
$\Omega(\a)$
as a function of the single complex structure modulus
$\alpha$ and its first four derivatives of the form
$\sum_{i=0}^4 f_i(\alpha)\Omega^{(i)}(\alpha)
={\rm d}\beta$. Integration of this relation over
an element of the third homology $H_3(M,\ZZ)$ gives the
Picard-Fuchs equation for the periods $\varpi(\alpha)$.

As shown in refs \cite{Griffiths} (see also: \cite{LSW} \cite{LT}
\cite{Candelas})
the periods can be written as
\be
\varpi_i=\int_{\gamma_1}\int_{\gamma_2}
\int_{\Gamma_i} {\omega\over p_1 p_2}
\label{30form}
\ee
where
$$
\omega=\sum_{i=1}^m (-1)^i x_i dx_1\wedge\dots\wedge
\widehat{dx_i}\wedge\dots\wedge dx_m
$$
$\Gamma_i$ is an element of $H_3(X,\ZZ)$ and $\gamma_i$
are small curves around $p_i=0$ in the $m$-dimensional
embedding space.

For the derivation of the Picard-Fuchs equation it is
crucial to note that ${\partial\over\partial x_i}
\left({f\over p_1^m p_2^n}\right)\omega$ is exact,
which leads to the following partial integration rule:
$$
{f {\partial p_1\over\partial x_i}\over p_1^m p_2^n}
={1\over m\!-\!1}{{\partial f\over x_i}\over p_1^{m\!-\!1} p_2^n}
-{n\over m\!-\!1}{f{\partial p_2\over\partial x_i}\over
p_1^{m-1}p_2^{n+1}}
$$
We have used these rules to compute the Picard-Fuchs
equations for those models for which we could construct
the mirrors.

Let us demonstrate this on the example of the
$\ZZ_2$ torus given by two quadrics $X_{(2,2)}$ in
$\IP^3(1,1,1,1)$:
\be
p_1={1\over 2}(x_1^2+x_2^2-2\, \alpha\,  x_3 x_4)=0,\qquad
p_2={1\over 2}(x_3^2+x_4^2-2\, \alpha\,  x_1 x_2)=0.
\ee
We use the notation
$$
\int {x_1^{i_1}\cdots x_4^{i_4}\over p_1^m p_2^n}=
\sym{i_1}{i_2}{i_3}{i_4}{m}{n}.
$$
We can then integrate by parts with respect to, say, $x_1$,
either by writing $x_1=\partial p_1/\partial x_1$
or using $x_2=-{1\over\alpha}\partial p_2/\partial x_1$.
This leads to the following partial integration rules (with similar
rules for partial integration w.r.t. $x_{2,3,4}$):
$$
\ba{rl}
\sym{i_1}{i_2}{i_3}{i_4}{m}{n}
&={i_1-1\over m-1}\sym{i_1-2}{i_2}{i_3}{i_4}{m-1}{n}
+{\alpha n\over m-1}\sym{i_1-1}{i_2+1}{i_3}{i_4}{m-1}{n+1}
\hfil (a)
\\ [4mm]
&=-{i_1\over\alpha(n-1)}\sym{i_1-1}{i_2-1}{i_3}{i_4}{m}{n-1}
+{m\over\alpha (n-1)}\sym{i_1+1}{i_2-1}{i_3}{i_4}{m+1}{n-1}\hfil (b)
\\ [4mm]
\ea
$$
One then finds
$$
\ba{rl}
\sym{2}{2}{0}{0}{1}{3}
&=-{1\over\a}\sym{1}{1}{0}{0}{1}{2}
+{1\over4\a}\left\{\sym{3}{1}{0}{0}{2}{2}
 +\sym{1}{3}{0}{0}{2}{2}\right\} \\ [4mm]
&=-{1\over2\a}\sym{1}{1}{0}{0}{1}{2}
 +{1\over2}\sym{1}{1}{1}{1}{2}{2}, \\ [4mm]
\ea
$$
where in the first step we have used rule (b) twice once for $x_1$ and
once for $x_2$; to obtain the second line we have substituted
$x_1^2+x_2^2=2 p_1+\alpha x_3 x_4$.
This leads to $\varpi^{\prime\prime}=4\sym{1}{1}{1}{1}{2}{2}
-{1\over\a}\varpi^\prime$.
By repeated use of the partial integration rules, we
can express $\sym{1}{1}{1}{1}{2}{2}$
in terms of $\varpi$ and $\varpi^\prime$ and
finally arrive at the period equation for the
$\ZZ_2$ torus
$$
\a(1-\a^4)\varpi^{\prime\prime}+(1-5\a^4)\varpi^\prime
-4\a^3\varpi=0.
$$
In the same way we have derived the following
differential equations for the periods of the two
models under investigation:
$$
\a^3(1-\a^8)\varpi^{(iv)}-2\a^2(1+7\a^8)\varpi^{\prime\prime\prime}
-\a(1+55\a^8)\varpi^{\prime\prime}+(9-65\a^8)\varpi^\prime
-16\a^7\varpi=0
$$
and
$$
\a^3(1-\a^{12})\varpi^{(iv)}
-2\a^2(5+7\a^{12})\varpi^{\prime\prime\prime}
+\a(23-55\a^{12})\varpi^{\prime\prime}+(49-65\a^{12})\varpi^\prime
-16\a^{11}\varpi=0
$$
The equations have regular singular points at
$\a=0$, $\a^{d_1+d_2}=1$ and $\a+\infty$. The solutions of
the indicial equations are $\{0_2,4_2\}$, $\{0,1_2,2\}$
and $\{2_4\}$ respectively, for our first model and
$\{0_2,8_2\}$, $\{0,1_2,2\}$ and $\{2_4\}$ for our second model
(the subscripts denote multiplicities).
We note that in contrast to the models considered in
\cite{Morrison} \cite{KT1} \cite{Font} there are two logarithmic
solutions at $\a=0$.

In terms of the variable $z=\alpha^{-(d_1+d_2)}$ these
equations may be rewritten in the form
$$
((\Theta^4-z(\Theta+a_1)(\Theta+a_2)(\Theta+a_3)(\Theta+a_4))
\tilde\varpi=0
$$
where $\Theta=z{d\over dz}$ and $\tilde\varpi=\alpha^2\varpi$.
The two cases correspond to the parameters
$\lbrace a_i\rbrace=\lbrace {1\over4},{3\over4},{1\over4},{3\over4}
\rbrace$ and $\lbrace {1\over6},{5\over6},{1\over6},{5\over6}\rbrace$,
respectively.
We will drop the tilde in $\tilde\varpi$ in the following.

\sect{The Yukawa couplings and the instanton numbers}

The simplest way to arrive at the mirror map, the Yukawa
couplings and the number of instantons is to follow
\cite{Morrison} (see also \cite{LT}).
This requires merely the knowledge of two
solutions of the period equation
in the neighbourhood of the singular point
$\a=\infty\,(\sim z=0)$, namely the pure power series
solution $\varpi_0(z)$
and the solution with one power of $\log z$, $\varpi_1(z)$.

If we normalize $\varpi_0(z)$ as $\varpi_0(z)=1+O(z)$, it is given as
$$
\varpi_0(z)=\sum_{n=0}^\infty\prod_{i=1}^4{(a_i)_n\over n!}z^n
\equiv\sum_{n=0}^\infty{\prod_{i=1}^k (d_i n)!\over
\prod_{j=1}^ m (w_j n)!}(\gamma^{-1}z)^n
$$
where $k$ is the number of polynomial constraints
(for our threefolds, $d-k=4$) and $\gamma=2^{12}$ and
$\gamma=2^8 3^6$ for the two models, respectively.
Notice that this solution may be represented as the multiple
contour integral $\oint {dx_1\ldots dx_m\over p_1 p_2}$, where we
expand the integrand for $\alpha\rightarrow \infty$. This corresponds
to the explicit evaluation of (\ref{30form}) with a judicious
choice for the cycle\footnote{We would like to thank P. Candelas
for discussions on this point and for providing us with a
preliminary version of \cite{CDP}.}.

We now introduce the
variable $x=z/\gamma$ and normalize $\varpi_1$ such
that
$$
t(x)\equiv{\varpi_1(x)\over\varpi_0(x)}\sim\log x\qquad{\rm for}
\qquad x\to 0
$$
This relation describes the mirror map. More explicitly,
if we make the ansatz
$\varpi_1(x)=\sum_{n=0}^\infty d_n x^n+c\varpi_0(x)\log x$
we find $c=1$, and $d_0=0$, and for the $d_n, n>0$
a recursion relation. The fully instanton corrected
Yukawa coupling on the manifolds $X^\prime$ is then
$$
\kappa_{ttt}=-\kappa_{ttt}^0{\varpi_0^4\over W^3(\gamma x(q)-1)}
$$
where $W=\varpi_0\Theta\varpi_1-\varpi_1\Theta\varpi_0$,
$x(q)$ is the inverse function of $q(x)\equiv\exp(t(x))$
and $\kappa_{ttt}^0$, the infinite radius limit of the
Yukawa coupling, is the intersection number
$\kappa_{ttt}^0=\int_M J^3=\prod_{i=1}^kd_i/\prod_{j=1}^d w_j$
($\{4;1\}$ in our examples).

As conjectured in \cite{CDGP} and proven in \cite{AM},
$\kappa_{ttt}$ can be expanded as
$$
\kappa_{ttt}=\kappa_{ttt}^0+\sum_{d=1}^\infty{n_d d^3 q^d\over
1-q^d}
$$
where $n_d$ denotes the number of rational curves of degree $d$.
(The denominator arises from summing over all multiple covers.)

With the given information it is now straightforward to compute
the $n_d$. We have listed the first few in Table 2.

\sect{Discussion}
There are three models left in Table 1. For those we have not
been able to construct the mirror by finding a symmetry group
$G$ such that the mirror manifold $X^\prime$ is given as
$X^\prime=\widehat{X/G}$.
To get information about the mirrors of these models
we may, inspired by the success of ref. \cite{LT},
extrapolate our knowledge of the parameters $a_i$ of
the period equation to these models. This leads to the following
choices for the three models in turn\footnote{
Note that in all cases considered here and in \cite{CDGP}\cite{Morrison}
\cite{KT1}\cite{Font}\cite{LT}
the $a_i$ are determined by $ \{a_i\}=\{
{\alpha\over d_p} | 0 < {\alpha\over d_p} < 1,
{\alpha\over d_p} \neq {\beta\over w_q},
\forall\, p=1,\ldots,k,\, q=1,\ldots,m,\, \alpha,\beta \in \IN\} $} :
$\{a_i\}=\{{1\over3},{2\over3},{1\over4},{3\over4}\}$,
$\{{1\over6},{3\over6},{5\over6},{1\over2}\}$,
$\{{1\over6},{5\over6},{1\over4},{3\over4}\}$.
The corresponding values for $\gamma$ are
$\gamma=\{3^3 2^6; 3^3 2^8;3^3 2^{10}\}$.
With $\kappa_{ttt}^0=\{6;4;2\}$
we do indeed find integer $n_d$'s.
The first few are listed
in Table 2. We thus conjecture that with these choices for
the parameters we do correctly describe the periods and
thus the Yukawa couplings of the mirror manifolds, even though
we do not know an explicit description for them.
$$\vbox{
{{
{\offinterlineskip\tabskip=0pt
\halign{\strut\vrule#&$#$~&
\vrule$#$&~\hfil$#$~&\vrule$#$&~\hfil$#$~&\vrule$#$\cr
\noalign{{\hrule}}
& &&\hfil N^0=1\quad \hfil && \hfil N^0=2\qquad\hfil &\cr
\noalign{\hrule}
&~n_0&&
4&&
1&\cr
&~n_1 && 3712
&&  67104
&\cr
&~n_2 && 982464
&& 847288224
&\cr
&~n_3 && 683478144
&& 28583248229280
&\cr
&~n_4 && 699999511744
&&  1431885139218998016
&\cr
\noalign{{\hrule}}
& &&\hfil N^0=3 \quad\hfil && \hfil N^0=4 \qquad\hfil &\cr
\noalign{\hrule}
&~n_0&&6&&4&\cr
&~n_1
&& 1944
&&  4992
&\cr
&~n_2
&&  223560
&& 2388768
&\cr
&~n_3
&&  64754568
&&  2732060032
&\cr
&~n_4
&&  27482893704
&&  4599616564224
&\cr
\noalign{{\hrule}}
& &&\hfil N^0=5 \quad\hfil && \hfil \qquad\hfil &\cr
\noalign{\hrule}
&~n_0&& 2
&&
&\cr
&~n_1
&&  15552
&&
&\cr
&~n_2
&&  27904176
&&
&\cr
&~n_3
&&  133884554688
&&
&\cr
&~n_4
&&  950676829466832
&&
&\cr
\noalign{\hrule}}}}}
}$$
\centerline{{\bf Table 2:} The numbers of rational curves of low degree}

\vskip 0.5truecm

In closing, we want to remark that the method of \cite{BCOV}
can in principle be used to determine also the number of
elliptic curves on the manifolds. Unfortunately,
the results of \cite{BCOV} can, however, not be immeadiately applied here.
For the cases they consider, the manifolds, and thus the
index $F_1$, is regular at $\a=0$, which was used as an input.
In contrast to this, our manifolds cease to be transversal at
$\a=0$.


\begin{thebibliography}{99}

\bibitem{Dixon} L. Dixon, in {\em Superstrings, Unified Theories and
                Cosmology 1987}, G. Furlan et al, eds.,
                World Scientific 1988

\bibitem{LVW} W. Lerche, C. Vafa and N. Warner, \npb324(1989)427

\bibitem{MM} \mm

\bibitem{CDP} P. Cadelas, E. Derrik and L. Parkes, {\em
              Generalized Calabi-Yau Manifolds and the Mirror
              of a Rigid Manifold}, preprint UTTG-24-92

\bibitem{CDGP} P. Candelas, X. De la Ossa, P. Green and
               L. Parkes, \npb359(1991)21

\bibitem{Morrison} D. Morrison: {\em Picard-Fuchs Equations
                and Mirror Maps for Hypersurfaces},
                in \mm

\bibitem{KT1} A. Klemm and S. Theisen, \npb389(1993)153

\bibitem{Font} A. Font, \npb391(1993)358

\bibitem{DG} J. Distler and B. Greene, \npb309(1988)295

\bibitem{LT} A. Libgober and  J. Teitelbaum,
             {\em Duke Math. Jour., Int. Math. Res. Notices}
             {\bf 1} (1993) 29

\bibitem{DK} D. Dais and A. Klemm:
            {\em Generalized Landau-Ginzburg String Vacua and
            Their Associated Calabi-Yau Spaces}, in preparation

\bibitem{Ks} A. Klemm and R. Schimmrigk:
            {\em Landau-Ginzburg String Vacua},
            CERN-TH-6459/92,
            to appear in {\sl Nucl. Phys.} {\bf B}

\bibitem{ks}M. Kreuzer and H. Skarke, \npb388(1992)113

\bibitem{Fletcher} A.R. Fletcher: {\em Working with Weighted
              Complete Intersections}, Max-Planck-Institut Series
              MPI/89-35, Bonn (1989)



\bibitem{MOP} D. Markushevich, M. Olshanetsky, A. Perelomov,
               {\sl Commun. Math. Phys.} {\bf 111} (1987) 247

\bibitem{Oda} T. Oda  : {\em Convex Bodies and Algebraic Geometry,
              (An Introduction to the Theory of Toric Varieties}),
              {\sl Ergebnisse der Mathematik und ihrer Grenzgebiete},
              3.Folge, Bd. {\bf 15}, Springer-Verlag (1988)

\bibitem{Roan1} S.-S. Roan,
               {\sl Int. Jour. Math.}, Vol. {\bf 2} (1991)
               439

\bibitem{DHVW} L. Dixon, J. Harvey, C. Vafa and E. Witten,
               \npb261(1985)678, \npb274(1986)285


\bibitem{Griffiths} P. Griffiths, {\sl Ann. of Math.} {\bf 90}
                                  (1969) 460

\bibitem{LSW}  W. Lerche, D. Smit and N. Warner,
                \npb372(1992)87

\bibitem{Candelas} P. Candelas, \npb298(1988)458

\bibitem{AM} P. Aspinwall and D. Morrison, {\em Topological Field
             Theory and Rational Curves}, preprint DUK-M-91-12


\bibitem{BCOV} M. Bershadsky, S. Cecotti, H. Ooguri and C. Vafa,
               {\em Holomorphic Anomalies in Topological
               Field Theories}, preprint HUTP-93/A008,
               RIMS-915.


\end{thebibliography}
\end{document}